\documentclass[pss,fleqn]{w-art}
\usepackage{times}
\usepackage{w-thm}
\usepackage{bm}% bold math
\usepackage{epsfig}
\usepackage[]{graphicx}
\begin{document}
%%    The information for the title page will be placed between
%%    \begin{document} and \maketitle. The order of most entries
%%    is determined by the class file and can not be changed by
%%    rearranging them. The maketitle command follows after the
%%    absract.
%%
%%    Most of the following commands will be completed by the publisher.
%%
\DOIsuffix{theDOIsuffix}
%%
%% issueinfo for header and copyright line
\Volume{XX} \Issue{1} \Copyrightissue{01} \Month{01} \Year{2004}
%%
%%    First and last pagenumber of the article. If the option
%%    'autolastpage' is set (default) the second argument may be left empty.
\pagespan{1}{}
%%
%%    Dates will be filled in by the publisher. The 'reviseddate' and
%%    'dateposted' (Published online) entry may be left empty.
\Receiveddate{\sf zzz} \Reviseddate{\sf zzz} \Accepteddate{\sf
zzz} \Dateposted{\sf zzz}
%%
%%    Give a maximum of six PACS code in numerical order.

%% \pretitle{Editor's Choice}

%% We have a short and a long form for the title. The short form
%% (optional argument) goes into the running head.
\title[Correlation effects in EuS]{Correlation effects in the valence bands of
  ferromagnetic semiconductor EuS  }

%% Please do not enter footnotes or \inst{}-notes into the optional
%% argument of the author command. The optional argument will go into
%% the header.  If there is only one address the marker \inst{x} may be
%% omitted.

%% Information for the first author.
\author[A.Sharma]{A.Sharma\footnote{Corresponding
     author: e-mail: {\sf anand@physik.hu-berlin.de}}\inst{1}}
\address[\inst{1}]{Institut f\"ur Physik, Humboldt-Universit\"at
     zu Berlin, Newtonstr.15, 12489, Berlin, Germany.}
%%
%% Information for the second author
\author[W.Nolting]{W. Nolting\inst{1}}
%%\footnote{Second author footnote.} may be inserted after the name.
%%
%% Information for the third author
%%\author[M. Stutzmann]{Martin Stutzmann\inst{1,2}}
%%\footnote{Third author footnote.} may be inserted after the name.
%%\address[\inst{2}]{Walter-Schottky-Institut, Technische Universit\"at M\"unchen,
%%Am Coulombwall, 85748 Garching, Germany}
%%
%%    \dedicatory{This is a dedicatory.}
\begin{abstract}
We present a many body analysis of the multi-band Kondo lattice
model. The study is then combined with the first principles
TB-LMTO band structure calculations, in order to investigate the
temperature dependent correlation effects in the 3$\textit{p}$
valence bands of the ferromagnetic semiconductor EuS. Some of the
physical properties of interest like the quasi-particle density of
states (Q-DOS), spectral density (SD) and quasi-particle band
structure (Q-BS) are calculated and discussed. Therewith, we
propose a spin resolved ARPES of the valence bands of EuS to be
performed.
\end{abstract}
%% maketitle must follow the abstract.
\maketitle                   % Produces the title.

%% If there is not enough space inside the running head
%% for all authors including the title you may provide
%% the leftmark in one of the following three forms:

%% \renewcommand{\leftmark}
%% {First Author: A Short Title}

%% \renewcommand{\leftmark}
%% {First Author and Second Author: A Short Title}

\renewcommand{\leftmark}
{A. Sharma and W.Nolting : Correlation effects in the valence bands of ferromagnetic
  semiconductor EuS.}

\section{\label{sec:intro} INTRODUCTION}

\indent The europium chalcogenides EuX, with X=O,S,Se and Te, formed the
subject of many studies for more than forty years. They are
magnetic semiconductors and one of the early motivation was the
hope of using these materials as magneto-optical memories in the
computers and magneto-optical modulators. Moreover, the europium
monochalcogenides are face centered cubic materials that are ideal
examples of Heisenberg exchange model and thus formed good
exemplary substances for studying magnetism. In such a motivated
background experiments were performed to study their properties
and use them for practical purposes. Though apart from various
dopings and substitutions, it turned out that the ordering
temperature could not be raised up to room temperature. Also,
these materials were found difficult to prepare. But, despite such
difficulties, the studies on these materials attracted attention
from a more academic point of view as it highlighted many
interesting physical properties. For example \cite{wachter,mauger}, a
metal-insulator transition in Eu-rich EuO has been observed while
red-shift of the absorption edge, Faraday rotation, circular
dichroism and spin polarization explained magnetic and
magneto-optical effects. Such interesting properties have not only
provided the motivation for theoretical and experimental
investigations on these materials, but further studies on them
permit a better understanding on more complex systems like the
Diluted Magnetic Semiconductors (DMS) which are believed to be one
of the essential materials behind the technology called spintronics \cite{wolf,zutic}.\\
\indent The motivation behind such a study is to combine many body
theoretical analysis with bandstructure calculations for real
materials. The goal is to study the temperature dependent
electronic correlation effects in the valence bands of EuS. A
similar study has been carried out earlier for the conduction
bands of EuS \cite{mueller1} but under a different kind of a theoretical
approach. The chalcogenide, EuS, was under a lot of experimental
and theoretical investigation to study its magnetic and electronic
properties, starting from early 1960s which reported nuclear
magnetic resonance \cite{heller} till the latest being on EuS
thin films \cite{mueller2} while the list remains extensive.\\
\indent This paper is organized as follows. In
section \textbf{\ref{sec:th-mo-cal}}, we formulate the complete multi-band
Hamiltonian. A self energy ansatz, reliable for low carrier
densities, is used to solve the model Hamiltonian. The temperature
and spin dependent density of states (DOS) are presented for two
weakly hybridized bands in a simple cubic lattice with coupling
strength, hybridization and magnetization being the parameters for
the model study. The results are then compared with one of the
exact limiting cases of the model, namely the ferromagnetically
saturated semiconductor and are found to be coinciding. In
section \textbf{\ref{sec:recal-EuS}}, such a many body model is combined
with first principles T=0 band-structure calculation, tight
binding linear muffin-tin orbital (TB-LMTO), in order to calculate
some of the physical properties like quasi-particle density of
states (Q-DOS), spectral density (SD) and quasi-particle band
structure (Q-BS) for realistic ferromagnetic semiconducting
material EuS. The final section \textbf{\ref{sec:con}}, concludes with the
proposition of a spin resolved ARPES experiment to be performed in order to examine our results.\\

\section{\label{sec:th-mo-cal} THEORY AND MODEL CALCULATION}

\begin{SCfigure}[4][h]
\includegraphics[height=5.0cm,width=7.0cm]{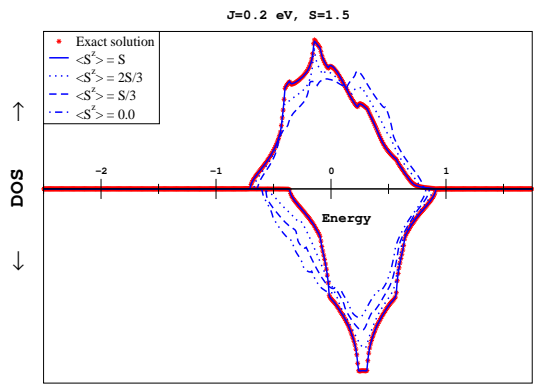}%
\caption{The spin up ($\sigma$=$\uparrow$)and down ($\sigma$=$\downarrow$) DOS, as function of energy, are calculated with J=0.2 eV, S=1.5 for various values of magnetization $\frac{\langle S^{z} \rangle}{S}$=1.0, 2/3, 1/3 and 0.0 represented as thick solid line, dotted line, broken line and double-dotted broken line, respectively. The case of ferromagnetically saturated semiconductor is represented as stars coinciding with $\frac{\langle S^{z} \rangle}{S}$=1.0.}
\label{fig:1}
\end{SCfigure}

\indent The multi-band Kondo lattice model (KLM) Hamiltonian reads as follows; \\
\begin{equation}\label{eq:Ham}
H=H_{kin}+H_{int}
\end{equation}
\indent where
\begin{equation}\label{eq:Ho}
H_{kin}=\sum_{ij\alpha\beta\sigma}T_{ij}^{\alpha\beta}c_{i\alpha\sigma}^{\dagger}c_{j\beta\sigma}
\end{equation}
\indent and
\begin{equation}\label{eq:Hint}
H_{int}=-\frac{J}{2}\sum_{i\alpha\sigma}(z_{\sigma}S_{i}^{z}c_{i\alpha\sigma}^{\dagger}c_{i\alpha\sigma}+S_{i}^{\sigma}c_{i\alpha-\sigma}^{\dagger}c_{i\alpha\sigma})
\end{equation}
\indent The term $H_{kin}$ denotes the kinetic energy of the
valence band electrons, 3$\textit{p}$ orbital in case of EuS, with
$c_{i\alpha\sigma}^{\dagger}$ and $c_{i\alpha\sigma}$ being the
fermionic creation and annihilation operators, respectively, at
lattice site $R_{i}$. The latin letters (i,j,...) as subscripts,
symbolize the crystal lattice indices while the band indices are
depicted as superscripts in Greek letters ($\alpha$,$\beta$,..)
and the spin is denoted as ${\sigma}(={\uparrow},{\downarrow})$.
Such a notation is used throughout the text. The multi-band
hopping term, $T_{ij}^{\alpha\beta}$, is connected by Fourier
transformation to the free Bloch energies $\epsilon^{\alpha\beta}$($\textbf{k}$)
\begin{equation}\label{eq:Hop}
T_{ij}^{\alpha\beta}=\frac{1}{N} \sum_{\textbf{k}}\epsilon^{\alpha\beta}(\textbf{k}) \hspace{0.2cm} e^{-i{\textbf{k}}\cdot (R_{i}-R_{j})}
\end{equation}
\indent $H_{int}$ is an intra-atomic exchange interaction term
being further split into two subterms. The first describes the
Ising type interaction between the z-component of the localized
and itinerant carrier spins while the other comprises spin
exchange processes which are responsible for many of the KLM
properties. J is the exchange coupling strength which we assume to
be $\textbf{k}$-independent and $S_{i}^{\sigma}$ refers to the
localized spin at site $R_{i}$
\begin{equation}\label{eq:Loc-spin}
S_{i}^{\sigma}=S_{i}^{x}+iz_{\sigma}S_{i}^{y} \hspace{0.2cm}; \hspace{0.2cm} z_{\uparrow}=+1, z_{\downarrow}=-1
\end{equation}
\indent The Hamiltonian in eq.~\eqref{eq:Ham} provokes a
nontrivial many body problem that cannot be solved exactly.
Approximations must be considered. We proceed to solve the problem
by the equation of motion method using double-time retarded Green
function \cite{zubarev}
\begin{equation}\label{eq:Green-fun}
G_{lm\sigma}^{\mu\nu}(E)=\langle\langle c_{l\mu\sigma};c_{m\nu\sigma}^\dagger\rangle\rangle_{E}
\end{equation}
where l,m and $\mu$,$\nu$ are the lattice and band indices
respectively. The equation of motion reads as follows
\begin{eqnarray}\label{eq:Eqn-mot}
EG_{lm\sigma}^{\mu\nu}(E) = \hbar \delta_{lm} \delta_{\mu\nu} + \sum_{j\gamma}
T_{lj}^{\mu\gamma}G_{jm\sigma}^{\gamma\nu}(E) -\frac{J}{2}[\Gamma_{lm\sigma}^{\mu\nu}(E)+F_{lm\sigma}^{\mu\nu}(E)]
\end{eqnarray}
where we use the following notations
\begin{subequations}
 \begin{equation}\label{eq:Hig-order-Is}
 \Gamma_{lm\sigma}^{\mu\nu}(E)=\langle\langle S_{l}^{z}c_{l\mu\sigma};c_{m\nu\sigma}^\dagger \rangle\rangle_{E}
 \end{equation}
\begin{equation}\label{eq:Hig-order-sf}
F_{lm\sigma}^{\mu\nu}(E)=\langle\langle S_{l}^{-\sigma}c_{l\mu-\sigma};c_{m\nu\sigma}^\dagger \rangle\rangle_{E} 
\end{equation}
\end{subequations}
\\
\indent We find that eq.(8a) and eq.(8b) are higher order Green
functions blocking the direct solution of the equation of motion
as they can't be decoupled into the forms of the original Green
function. But, a rather formal solution can be stated as
\begin{equation}\label{eq:Green-matrix}
\widehat{G}_{\textbf{k}\sigma}(E)=\hbar\widehat{I}[{(E+i0^{+})\widehat{I}-\widehat{\epsilon}(\textbf{k})-\widehat{\Sigma}_{\textbf{k}\sigma}(E)}]^{-1}
\end{equation}
where we exclude the band indices by representing the terms in a
generalized matrix form on symbolizing a hat over it
\begin{equation}\label{eq:Green-Fourier}
\widehat{G}_{lm\sigma}(E)=\frac{1}{N}\sum_{\textbf{k}}\widehat{G}_{\textbf{k}\sigma}(E)
\hspace{0.2cm} e^{-i{\textbf{k}} \cdot (R_{l}-R_{m})}
\end{equation}
\indent The terms in eq.~(\ref{eq:Green-matrix}) are explained as
follows : $\hat{I}$ is an identity matrix and $\hat{\epsilon}(\textbf{k})$ is a hopping matrix with the diagonal
terms of the matrix exemplifying the intra-band hopping and the
off-diagonal terms denoting the inter-band hopping. The self
energy, $\Sigma_{lm\sigma}^{\mu\nu}$(E), containing all the
influences of the different interactions being of fundamental
importance can be understood as
\begin{equation}\label{eq:self-energy}
\langle\langle[H_{int},c_{l\mu\sigma}]_{-};c_{m\nu\sigma}^{\dagger}\rangle\rangle=\sum_{p\gamma}\Sigma_{lp\sigma}^{\mu\gamma}(E)G_{pm\sigma}^{\gamma\nu}(E)
\end{equation}
\indent Now, we are left with a problem of finding a multi-band
self energy ansatz, in order to compute the Green function matrix
and thereby calculate some of the physical quantities of interest
like the quasi-particle spectral density (SD)
\begin{equation}\label{eq:SD}
A_{\textbf{k}\sigma}(E)=-\frac{1}{\pi}ImTr \widehat{G}_{\textbf{k}\sigma}(E)
\end{equation}
and the quasi-particle density of states (Q-DOS)
\begin{equation}\label{eq:DOS}
\rho_{\sigma}(E)=\frac{1}{N\hbar}\sum_{\textbf{k}}A_{\textbf{k}\sigma}(E)
\end{equation}
\indent As observed from our many body theoretical model, we are
only interested in the influence of inter-band exchange on the
valence band states in order to study the electronic correlations
and not aimed at calculating the magnetic properties via self
consistent calculation of the localized magnetization. For this
purpose, we neglect the exchange interaction between the localized
spins, $S_{i}^{\sigma}$, in our model. Furthermore, since we
consider fully occupied valence bands, the Coulomb interaction is
unimportant and this can further reduce the complexity in
formulating the self energy ansatz. Actually, in case of realistic
calculations of EuS, the Coulomb interaction is implicitly taken
in the LSDA bandstructure calculations. So, the single particle
excitation energies contain the Coulomb interactions and need not
be taken into account in the self energy ansatz. Thus, the
criterion required by our multi-band electronic self energy is
that it should be accurately defined in the low carrier density
regime and weak coupling strength, which is the case for the study
of magnetic semiconductors.\\
\indent It is known that the above statements can be satisfied by
the interpolating self energy ansatz (ISA) being proposed for a
single band model \cite{nolting1}. Such an ansatz is considered to work
for all coupling strengths and band occupations. Various exactly
known limiting cases like the weak and strong coupling behavior,
zero bandwidth limit \cite{nolting2}, etc. are shown to be satisfied by
the single-band ISA. Though it is not self-evident in case of
multi-band situation that these confining cases would hold true,
nevertheless we propose to replace the single particle Green
function for the multi-band case by the respective matrix Green
function. As seen in theoretical analysis given in Appendix~\ref{app:fss}, the
important case of ferromagnetically saturated semiconductor gives
the same self-energy ansatz as in the single band case (eq.(26) and
eq.(27) in Ref. 9) which strongly supports our ansatz. With the
help of such analysis, we put forward for the multi-band
self-energy (ISA);
\begin{equation}\label{eq:Mult-SE}
\widehat{\Sigma}_{\sigma}(E)=\frac{J}{2}M_{-\sigma}\widehat{I}+\frac{J^{2}}{4}a_{-\sigma}\widehat{G}_{-\sigma}(E+\frac{J}{2}M_{-\sigma})\left[\widehat{I}-\frac{J}{2}\widehat{G}_{-\sigma}(E+\frac{J}{2}M_{-\sigma})\right]^{-1}
\end{equation}
where
\begin{equation}
M_{\sigma}=z_{\sigma} \langle S^{z} \rangle; \hspace{0.2cm}a_{\sigma}=S(S+1)+M_{\sigma}(M_{\sigma}+1). \nonumber\\
\end{equation}
and the bare Green function matrix is defined as :
\begin{equation}\label{eq:Eff-Green}
\widehat{G}_{\sigma}(E)=\frac{1}{N}\sum_{\textbf{k}}\frac{1}{(E+i0^{+})\widehat{I}-\widehat{\epsilon}(\textbf{k})} \nonumber\\
\end{equation}
\indent It retains the formerly proposed ISA structure in the single band
case. As seen in eq.~\eqref{eq:Mult-SE}, we are interested only in
the local self energy
\begin{equation}\label{eq:SE-k-ind}
\widehat{\Sigma}_{\sigma}(E)=\frac{1}{N}\sum_{\textbf{k}}\widehat{\Sigma}_{\textbf{k}\sigma}(E)\nonumber\\
\end{equation}
\indent The physical reason for the wave-vector dependence of the
self energy is mainly due to the magnon energies
$\hbar\omega(\textbf{k})$ appearing at finite temperature. But, as
stated earlier, we neglect a direct Heisenberg exchange between
the localized spins and this can be interpreted as the
$\hbar\omega(\textbf{k})$ $\rightarrow$ 0 limit. The localized
magnetization $\langle S^{z} \rangle$ shall be considered as an
external parameter being responsible for the induced temperature
dependence of the band states.\\
\indent In principle, apart from the above discussed proposal one could also 
consider other alternative approaches for the many body analysis like the 
dynamical mean field theory (DMFT) in order to study the correlation effects. 
But, it is known \cite{georges} that the methods of DMFT can not be 
directly applied to the Kondo lattice model. One exception is the case 
of the classical spin limit, thus removing the quantum nature of the spins. 
While there is another possibility to derive a DMFT for Kondo lattice model, 
based on the fermionization of the localized spin operators as suggested 
in \cite{matsumoto}, but it is limited to S=$\frac{1}{2}$. Recently, in an 
article \cite{meyer}, the correlation effects were discussed by introducing 
DMFT based approaches but for a single band Kondo lattice model and with an 
evaluation compared with the exactly solvable limiting cases and several other 
known approximate methods.\\  
\indent Considering our multi-band electronic self energy ansatz, we proceed 
to perform a model calculation of two weakly hybridized bands on a
simple cubic (sc) lattice where the terms in the (2 x 2) hopping
matrix, $\widehat{\epsilon}(\textbf{k})$ , are taken using the sc
Bloch density of states (DOS) in the tight binding
approximation \cite{jelitto}. We assume the bandwidth (W) of both the
bands, i.e intra-band transfer energy to be 1.0 eV while that for
inter-band to be 0.10 eV. The center of gravity of one of the free
Bloch band is chosen to be energy zero while for the other it is taken as 0.25 eV. \\
\indent In order to study the electronic correlations, one
examines the effect of a test electron by creating (or
annihilating) it in an empty (or filled ) band. In our case, we
first annihilate an electron from the completely filled band.
Thus, removal of $\downarrow$($\uparrow$) electron will create a
$\uparrow$($\downarrow$) hole which will give insight into the
study. The correlation effects are studied by calculating the
electronic self energy. But, apart from electronic sub-system, we
also have the magnetic sub-system. The exchange coupling between
the itinerant electron and localized spins adds up to the
correlation effects as they produce spin-flip transitions and
Ising like interactions in addition to the kinetic energy. \\
\indent Fig.~\ref{fig:1} describes the weak coupling behavior,
JS$\ll$W. The calculated density of states (DOS) are shown for the
parameters : J=0.2 eV, S=1.5 and different values of
magnetization, $\frac{\langle S^{z} \rangle}{S}$=1.0, 2/3, 1/3 and
0.0 which are represented as thick solid line, dotted line, broken
line and double-dotted broken line, respectively. The exact
solution, ferromagnetically saturated semiconductor, is
represented as stars and it is found to coincide with results of
$\langle S^{z}\rangle$=S. 
\\\\
\\\\
\begin{figure}[h]
\begin{minipage}[t]{.45\textwidth}
\includegraphics[height=5.0cm,width=7.0cm]{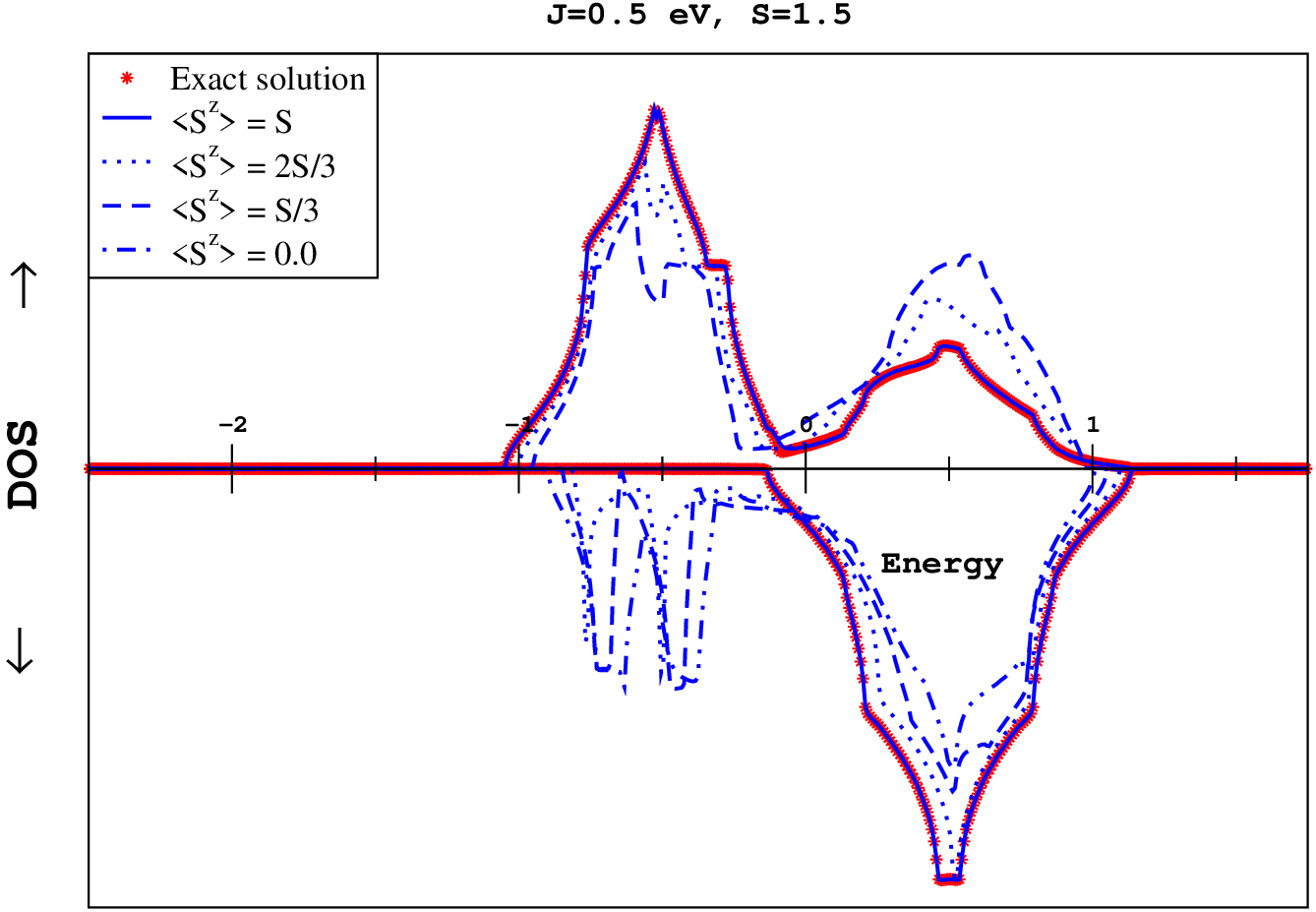}
\caption{The same as Fig.~\ref{fig:1} except for J=0.5.}
\label{fig:2}
\end{minipage}
\hfil
\begin{minipage}[t]{.45\textwidth}
\includegraphics[height=5.0cm,width=7.0cm]{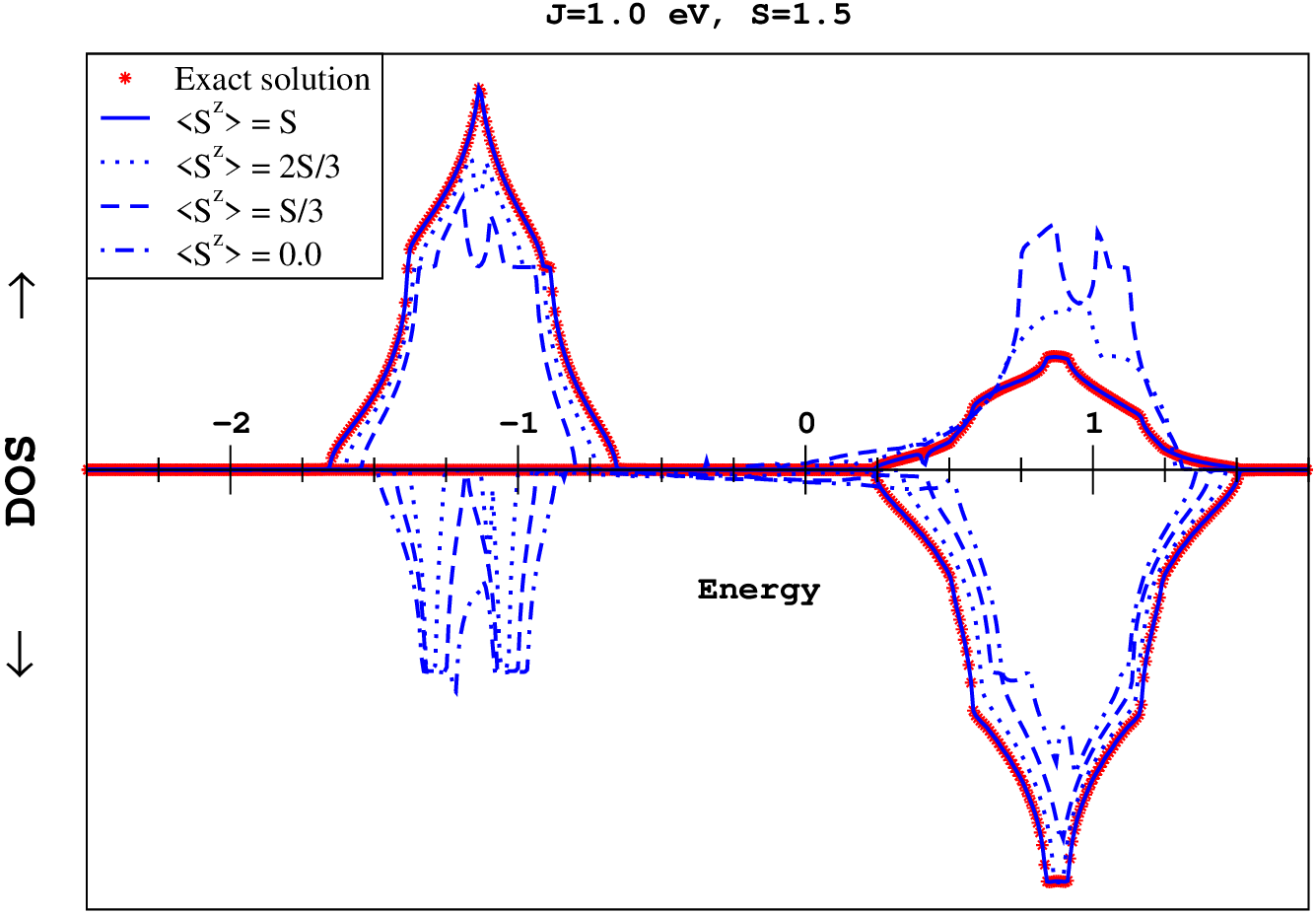}
\caption{The same as Fig.~\ref{fig:1} except for J=1.0.}
\label{fig:3}
\end{minipage}
\end{figure}
\\\\
\indent At T=0 K and fully occupied bands, the removal of a
$\downarrow$ electron (i.e. creation of a $\uparrow$ hole)
produces a stable quasi-particle since the $\uparrow$ hole has no
chance to flip its spin with the ferromagnetically saturated spin
($\Uparrow$) sub-system. The imaginary part of the self energy
vanishes indicating infinite lifetimes. The situation is different
for the removal of an $\uparrow$ electron (i.e, creation of
$\downarrow$ hole) as the $\downarrow$ hole can exchange its spin
with the ferromagnetically saturated spin sub-system and have
finite lifetime. It can emit a magnon and become a $\uparrow$ hole
provided there exist $\uparrow$ hole states which can be occupied
after spin-flip process. These are the scattering states which
occupy the same energy regions as of the $\uparrow$ spectrum. The
scattering states become more distinct with increasing coupling
strength as seen in Fig.~\ref{fig:2} and Fig.~\ref{fig:3}. The
other possibility for the $\downarrow$ hole to exchange its spin
could be by repeated emission and absorption of magnons, thus
forming a quasi-particle called magnetic polaron.\\
\indent However, at finite temperatures, the spin sub-system is no
longer perfectly aligned. There are magnons in the system that can
be absorbed by the itinerant charge carriers. As seen from the
figures, at finite temperatures the spectral weight gets
redistributed due to spin flip term in the exchange interaction
with deformations in the density of states. In the limit $T$
$\rightarrow$ $T_{c}$ ($\langle S^{z} \rangle \rightarrow 0$) the
induced spin asymmetry is removed.\\

\section{\label{sec:recal-EuS} PHYSICAL PROPERTIES : $\textbf{EuS}$}

\begin{SCfigure}[4][h]
\includegraphics[height=5.0cm,width=6.0cm]{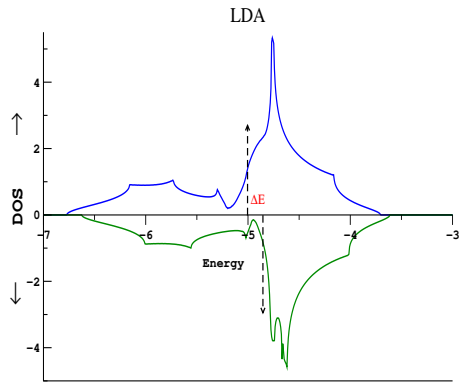}%
\caption{Spin-dependent density of states of 3$\textit{p}$ bands of EuS. Using the center of gravity of the bands, the exchange splitting amounts to $\Delta$E=0.1512 eV.}
\label{fig:cog}
\end{SCfigure}

\begin{figure}[htb]
\begin{minipage}[t]{.5\textwidth}
\includegraphics[height=5.0cm,width=7.0cm]{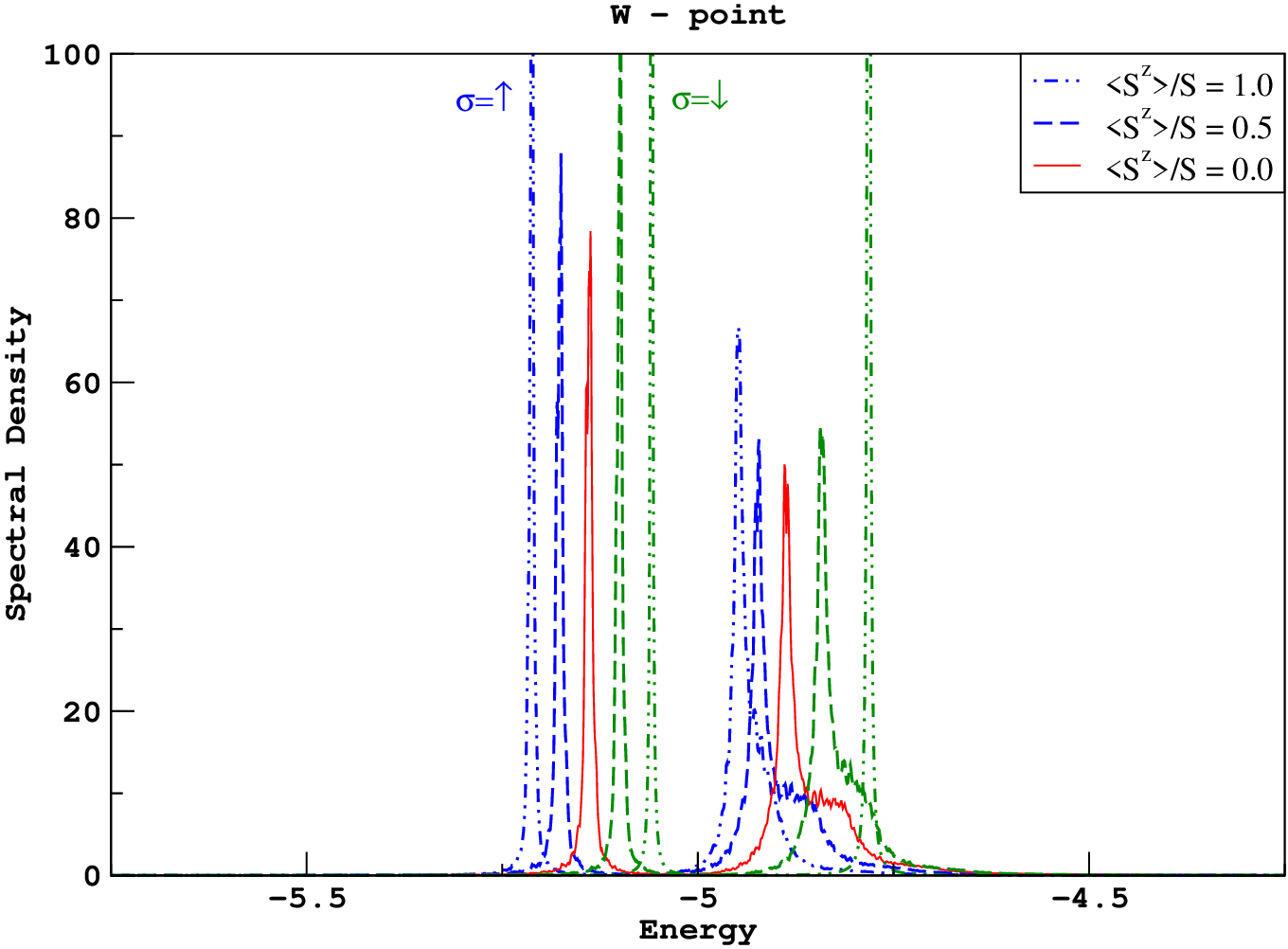}
\caption{\label{fig:w-pt} Spectral density at W - point.}
\end{minipage}
\hfil
\begin{minipage}[t]{.5\textwidth}
\includegraphics[height=5.0cm,width=7.0cm]{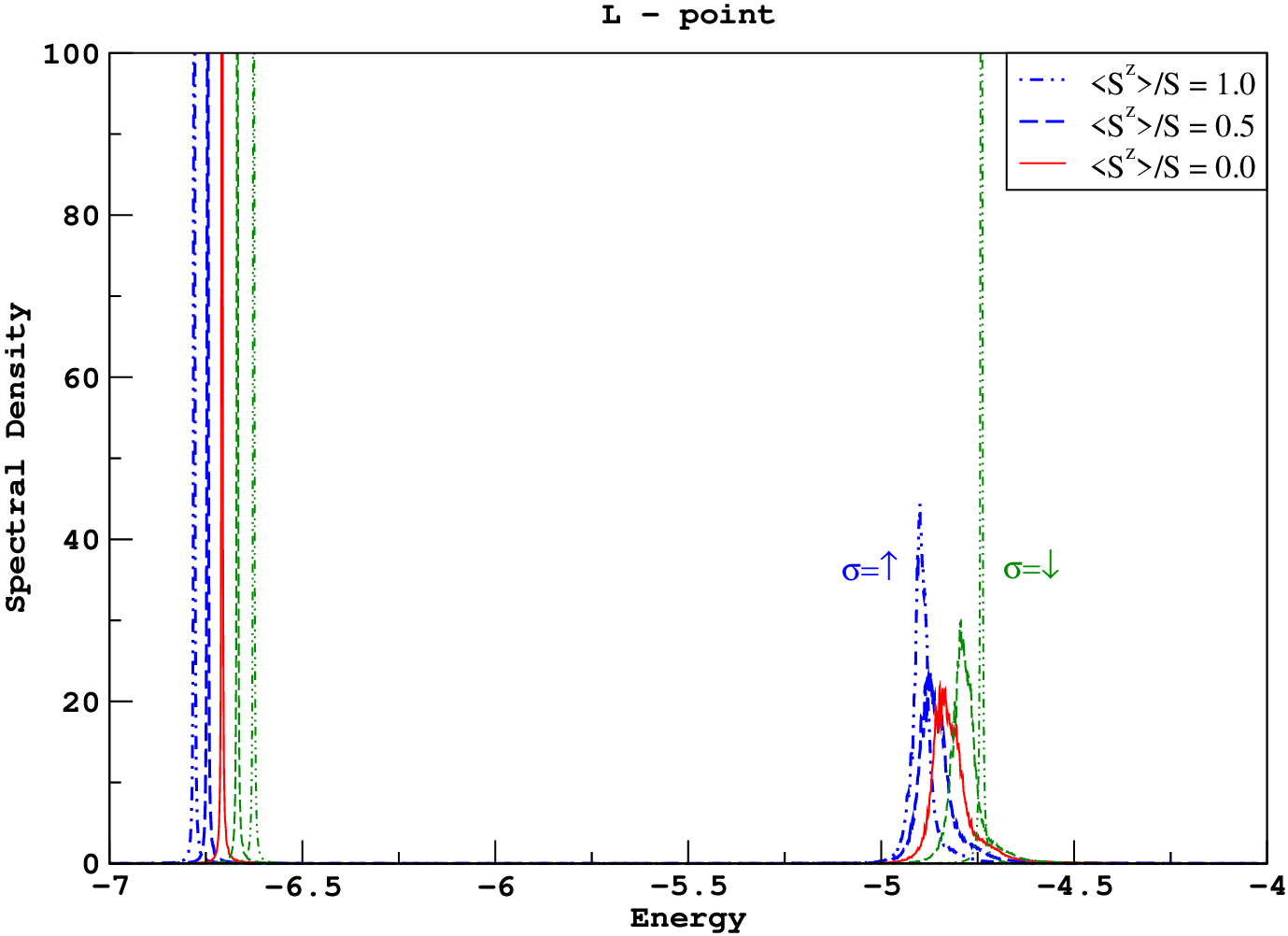}
\caption{\label{fig:l-pt} Spectral density at L - point.}
\end{minipage}
\end{figure}

\begin{figure}[htbp]
\centering
\includegraphics[height=5.0cm,width=8.0cm]{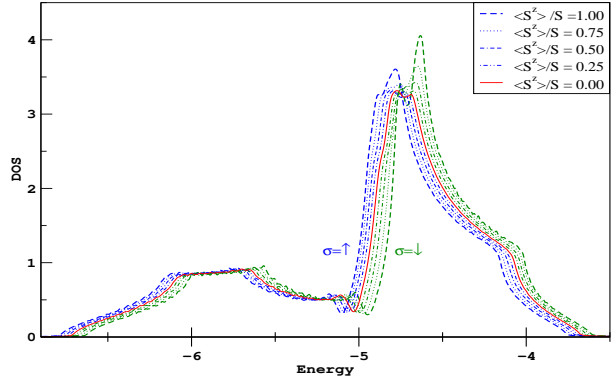}
\caption{\label{fig:qp-dos} Quasi-particle density of states
(Q-DOS) of the 3$\textit{p}$ valence bands of EuS.}
\end{figure}

\indent In this section, we study the electronic correlations in
the 3$\textit{p}$ valence bands of ferromagnetic semiconductor EuS
by combining the multi-band self energy ansatz for the many body
part along with the first principles TB-LMTO bandstructure
calculations.\\
\indent EuS crystallizes in a rocksalt structure with lattice
constant, a=5.95 $\AA$. Each $Eu^{2+}$ ion has twelve nearest and
six next nearest Eu-neighbors and they occupy lattice sites of a
fcc structure. The magnetism in EuS is due to the half-filled
4$\textit{f}$ shell of $Eu^{2+}$ and its magnetic properties are
well described using the Heisenberg model with the ferromagnetic
transition temperature, $T_{c}$=16.54 K. In this material, the
nearest neighbor exchange interaction is ferromagnetic while the
next nearest neighbor is anti-ferromagnetic.\\
\indent In order to have the single particle excitation energies,
i.e the hopping matrices, and all the interactions which are not
directly included in our Hamiltonian, we perform the TB-LMTO
bandstructure calculations using the program of
Anderson \cite{anderson1,anderson2}.
\\
\begin{SCfigure}[4][h]
\includegraphics[height=4.0cm,width=5.0cm]{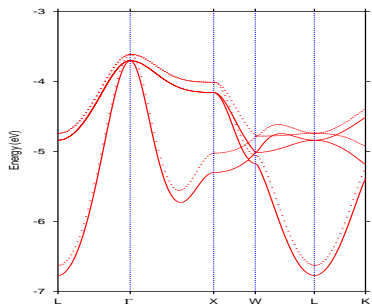}%
\caption{ LDA bandstructure of $3\textit{p}$ valence bands of EuS. Full lines are spin $\uparrow$ and dotted ones are spin $\downarrow$.}
\label{fig:lda-bndstr}
\end{SCfigure}
\indent Fig.~\ref{fig:lda-bndstr} shows the spin-dependent
bandstructure (LSDA) of the valence bands of EuS. In TB-LMTO, the
original Hamlitonian is transformed to a tight-binding Hamlitonian
containing only the nearest neighbor correlations. The evaluation
is restricted only to 3$\textit{p}$ bands. There are some
difficulties which arise due to the strongly localized character
of 4$\textit{f}$ levels and in order to avoid such difficulties,
we consider the seven 4$\textit{f}$ electrons as core electrons,
since these electrons enter our model only as localized moments.
\\
\indent The LDA-density of states is displayed in
Fig.~\ref{fig:cog}. Assuming that the LDA treatment of the
ferromagnetism is quite compatible with the Stoner (mean field)
picture, the T=0 splitting amounts to $\Delta$E=JS. On taking the
centers of gravity of the DOS for both the spins and along with
the above assumption, the exchange splitting amounts to
$\Delta$E=0.1512 eV which results in the value for the exchange
coupling strength as J=0.0432 eV where the spin value, S=3.5. One
can also calculate the exchange coupling by taking the energy
difference of the upper edge shift between both the spins which
gives J=0.0253 eV or calculate the mean value of the spin
polarized energy difference taken along each k-points in the first
Brillouin zone which gives J=0.0529 eV. It was found that the
coupling strength obtained by center of gravity approach, reflects
more correlation effects as was shown in the previous
work \cite{mueller1} for the conduction bands of EuS.
\\
\indent The single particle output obtained from the bandstructure
calculations are in the form of Hamiltonian and overlap matrix,
posing a generalized eigenvalue problem to be solved. In order to
employ such matrices as an input for the many body calculations,
one has to perform a Cholesky decomposition so as to reduce the
generalized problem to an eigenvalue problem. Using such an
approach, one can factorize the Hamiltonian matrices and then
obtain new Hamiltonian matrices using overlap matrices, which can
be used directly as hopping integrals,
$\widehat{\epsilon}(\textbf{k})$. We use the multi-band self
energy ansatz eq.~(\ref{eq:Mult-SE}) to compute the Green function
eq.~(\ref{eq:Green-fun}) and therewith calculate the spectral
densities eq.~(\ref{eq:SD}) and densities of states eq.~(\ref{eq:DOS}).
\\
%\DeclareGraphicsRule{.eps.gz}{eps}{.eps.bb}{'gunzip -c #1}
\begin{figure}[h]
\centering
\includegraphics[height=10.0cm,width=9.0cm]{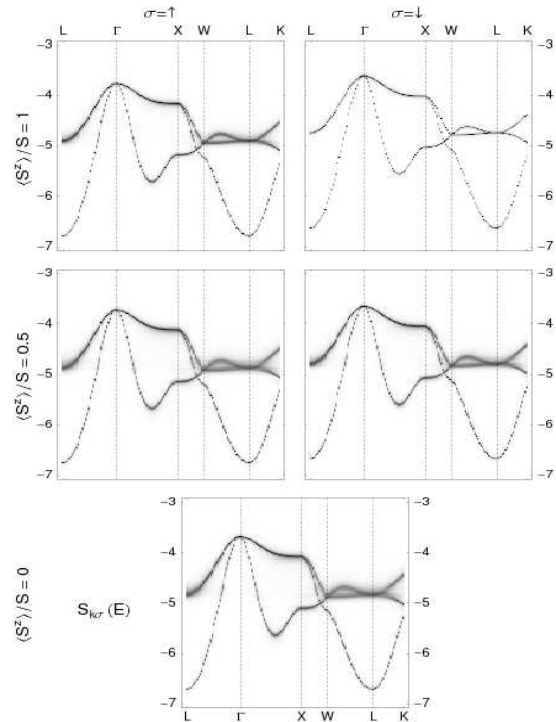}
\caption{\label{fig:qp-bndstr} Spin-dependent quasi-particle
bandstructure of 3$\textit{p}$ valence bands of EuS for different
values of magnetization $\frac{\langle S^{z} \rangle}{S}$.}
\end{figure}
\indent Fig.~\ref{fig:qp-bndstr} represents the quasi-particle
band-structure for some high-symmetry directions in the first
Brillouin zone. The degree of blackening measures the magnitude of
the spectral function. Correlation effects are clearly visible
along many parts, especially along the center of the valence band.
As seen in the $\uparrow$ spectrum even at T=0 K ($\frac{\langle
S^{z} \rangle}{S}$=1), parts of the dispersions are washed out
showing lifetime effects due to correlation in terms of magnon
emission and re-absorption with simultaneous spin-flips as
explained earlier. But at the same temperature, the $\downarrow$
spectrum shows sharp lines indicating infinite lifetime as
expected. While at finite temperature ($\frac{\langle S^{z}
\rangle}{S}$=0.5), the finite lifetimes appear in case of both the
spin spectra as the ferromagnetically saturated spin system is not
perfectly aligned giving rise to magnons which can be absorbed by
the itinerant charge carriers. At transition temperature
($\frac{\langle S^{z} \rangle}{S}$=0.0), the spin asymmetry is
removed. Strong temperature dependent correlation effects are seen
mainly along the W-L directions. \\
\indent In order to have a closer look at the correlations, the
spin and $\textbf{k}$-dependent spectral densities are plotted at
some high-symmetry points (W,L,X) in the first Brillouin zone. It
can be compared with the angle-resolved photoemission experiments
and for this purpose we propose that an experiment be performed. \\
\vspace{0.5cm}
\begin{figure}[htbp]
\centering
\includegraphics[height=5.0cm,width=7.0cm]{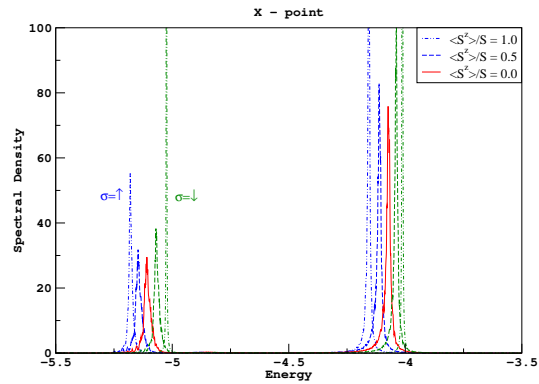}
\caption{\label{fig:x-pt} Spectral density at X - point.}
\end{figure}
\indent As seen in Fig.~\ref{fig:x-pt}, well defined
quasi-particle peaks appear with an additional spin split below
$T_{c}$. The exchange splitting collapses for $T \rightarrow
T_{c}$ and the quasi-particle damping increases with increasing
temperature. Similar explanations hold for the other high-symmetry
points (Fig.~\ref{fig:w-pt} and Fig.~\ref{fig:l-pt}). The
correlations are clearly observed at the W-point. With increasing
temperature, there's a strong damping as seen along the energy
spectrum. Interestingly, the same W-point remains the point of
discussion in case of the conduction band calculations of
EuS \cite{mueller1}. \\
\\\\\\
\indent The quasi-particle density of states (Q-DOS) is plotted in
Fig.~\ref{fig:qp-dos}. As observed, due to weaker exchange
coupling, the appearance of polaronlike quasi-particle branches
are less likely in the real material calculation. The temperature
influence on the spectrum is seen in terms of strong deformations,
mostly at the center, and in terms of shifts. While the
$\textbf{k}$-dependent spectral densities refers to the angle
resolved photoemission, the Q-DOS can be considered as the angle
integrated part. \\

\section{\label{sec:con} SUMMARY and CONCLUSION}

\indent In this paper, we have studied the effects of electronic
correlation on the 3$\textit{p}$ valence bands of the
ferromagnetic semiconductor EuS. We started with multi-band Kondo
lattice Hamiltonian which is sufficient enough to explain the
necessary physics. Using Green function approach, we try to
evaluate the Hamiltonian. We put forward an Interpolating Self
energy Ansatz (ISA) for the multi-band situation and obtain the
solution, i.e the required Green function, to predict some
physical properties. In order to test the theory, we perform model
calculations and interesting features are highlighted. For two
weakly hybridized bands assuming free Bloch densities of states on
a simple cubic lattice, polaron-like and scattering states are
observed for intermediate and strong coupling strengths. Strong
temperature dependent correlation effects are noticed in the form
of deformations and shifts in the energy spectrum. A non-trivial
exactly solvable limiting case of the model is correctly
reproduced thus confirming the proposed multi-band self energy ansatz.\\
\indent Such a self energy ansatz, which amounts to the many body
interactions, is later on combined with the first principles
TB-LMTO bandstructure calculations for calculating the properties
of real materials. The bandstructure output in the form of
hamiltonian and overlap matrices of the 3$\textit{p}$ valence band
electrons of EuS, serve as an input in the form of hopping
integrals for the many body calculations. The exchange coupling
strength is obtained from the LDA bandstructure calculations as a
result of energy difference in the center of gravities of the spin
polarized density of states. With these parameter inputs, physical
properties like quasi-particle spectral densities (Q-SD) and
densities of states (Q-DOS) are estimated for different values of
magnetization. One can also obtain other interesting physical 
quantities like the lifetime or the effective mass of the quasi-particles 
but it is not planned for the present article. While, in one 
of the previous work \cite{metzke}, emphasis was laid upon calculating
the spin flip probabilities which exhibit the lifetime effects. These 
quantities were calculated using the time-dependent spectral density 
which were used to express the spin polarization of the field emitted 
electrons thus highlighting the issue of spin filter effects of the 
europium chalcogenides. Eventually, as a consequence of our present 
calculations, the temperature dependent correlation effects are 
observed at the center of band and mainly in the W-L direction. 
For the sake of comparison, we propose and spin-dependent ARPES to 
be performed on the valence bands of EuS. \\

\begin{acknowledgement}
\indent One of us (A.S) would like to thank Dr.W.M\"uller for his
introduction to the TB-LMTO program and for helpful discussions.
\end{acknowledgement}

\appendix
\section{Case of ferromagnetically saturated semiconductor}
\label{app:fss}

\indent In the main text of this paper, we mentioned one limiting
approach of the multi-band model namely the ferromagnetically
saturated semiconductor which exhibits the quasi-particle called
magnetic polaron. It is that quasi-particle where a bare electron
(or hole) is dressed by a virtual cloud of magnons. In this
section, we present the analytical result of it. We are interested
in finding a solution for the model at T=0 with ferromagnetically
saturated spin sub-system ($\langle S^{z} \rangle$ = S) and for fully
occupied bands and $\sigma$=$\uparrow$,$\downarrow$. \\
\indent In such situation, the Ising-like higher order Green
function eq.~\eqref{eq:Hig-order-Is} becomes;
\begin{equation}\label{eq:Ising-mp}
\Gamma_{lm\sigma}^{\mu\nu}(E)= SG_{lm\sigma}^{\mu\nu}(E)
\end{equation}
\indent Now, let us consider : a) $\sigma$ = $\uparrow$ and b)
$\sigma$ = $\downarrow$ and try to evaluate the self energy in both the cases. \\
\indent a) For $\sigma$ = $\uparrow$, the equation of motion for
the spin-flip function
\begin{equation}
F_{lm\sigma}^{\mu\nu}(E)=\langle\langle S_{l}^{-\sigma}c_{l\mu-\sigma};c_{m\nu\sigma}^\dagger \rangle\rangle_{E} \nonumber\\
\end{equation}
\indent reads as follows,
\begin{equation}
EF_{lm\downarrow}^{\mu\nu}(E)=\langle\langle[S_{l}^{-\sigma} c_{l\mu\downarrow},H]_{-};c_{m\nu\uparrow}^{\dagger} \rangle\rangle_{E} \nonumber\\
\end{equation}
\indent which takes the form as given below yielding higher order
Green functions,
\begin{multline}
EF_{lm\sigma}^{\mu\nu}(E)=\sum_{p\gamma}T_{lp}^{\mu\gamma}\langle\langle S_{l}^{-} c_{p\gamma\downarrow};c_{m\nu\uparrow}^\dagger \rangle\rangle_{E}
-\frac{J}{2}[\langle\langle S_{l}^{-} S_{l}^{+} c_{l\mu\uparrow};c_{m\nu\uparrow}^\dagger \rangle\rangle_{E} - \langle\langle S_{l}^{-} S_{l}^{z}c_{l\mu\downarrow};c_{m\nu\uparrow}^\dagger \rangle\rangle_{E} \nonumber\\
+ \sum_{\lambda}\{(\sum_{\sigma}\langle\langle S_{l}^{-}
c_{l\lambda\sigma}^{\dagger}c_{l\lambda\sigma}c_{l\mu\downarrow};c_{m\nu\uparrow}^\dagger
\rangle\rangle_{E}) - 2 \langle\langle S_{l}^{z}c_{l\lambda\downarrow}^{\dagger}c_{l\lambda\uparrow}c_{l\mu\downarrow};c_{m\nu\uparrow}^\dagger
\rangle\rangle_{E}\} ] \\
\end{multline}
\indent The higher order Green functions resulting from the
spin-flip equation of motion can be evaluated exactly for the case
of fully occupied bands and at T=0,
\begin{subequations}\label{eq:spdecouple}
\begin{equation}\label{eq:a1}
\langle\langle S_{l}^{-} S_{l}^{+}
c_{l\mu\uparrow};c_{m\nu\uparrow}^\dagger \rangle\rangle_{E} = 0
\end{equation}
\begin{equation}\label{eq:a2}
\langle\langle S_{l}^{-} S_{l}^{z}
c_{l\mu\downarrow};c_{m\nu\uparrow}^\dagger \rangle\rangle_{E} =
SF_{lm\uparrow}^{\mu\nu}(E) \\
\end{equation}
\begin{equation}\label{eq:a3}
\langle\langle S_{l}^{-}(
c_{l\lambda\uparrow}^{\dagger}c_{l\lambda\uparrow} -
c_{l\lambda\downarrow}^{\dagger}c_{l\lambda\downarrow} )
c_{l\mu\downarrow};c_{m\nu\uparrow}^\dagger \rangle\rangle_{E} =
\delta_{\mu\lambda}F_{lm\uparrow}^{\lambda\nu}(E) \\
\end{equation}
\begin{equation}\label{eq:a4}
\langle\langle S_{l}^{z}
c_{l\lambda\downarrow}^{\dagger}c_{l\lambda\uparrow}c_{l\mu\downarrow};c_{m\nu\uparrow}^\dagger
\rangle\rangle_{E} = - \delta_{\mu\lambda}
SG_{lm\uparrow}^{\lambda\mu}(E) \\
\end{equation}
\end{subequations}
\indent On substituting the eq.~\eqref{eq:spdecouple} in the
spin-flip equation of motion and taking the Fourier transform we
get;
\begin{equation}
(E-\frac{JS}{2})F_{\textbf{k},\textbf{k-q},\textbf{q}\uparrow}^{\mu\nu}(E)
= \sum_{\lambda}\epsilon^{\mu\lambda}(\textbf{k-q})F_{\textbf{k},\textbf{k-q},\textbf{q}\uparrow}^{\lambda\nu}(E) - \frac{J}{2N}\sum_{\textbf{t}}F_{\textbf{k},\textbf{k-t},\textbf{t}\uparrow}^{\mu\nu}(E)-\frac{JS}{\sqrt{N}}G_{\textbf{k}\uparrow}^{\mu\nu}(E)\nonumber
\end{equation}
\indent and in the matrix form, it can be written as follows
\begin{equation}
[(E-\frac{JS}{2})\widehat{I}-\widehat{\epsilon}(\textbf{k-q})]\widehat{F}_{\textbf{k},\textbf{k-q},\textbf{q}\uparrow}(E) = - \frac{J}{2N}\sum_{\textbf{t}}\widehat{F}_{\textbf{k},\textbf{k-t},\textbf{t}\uparrow}(E)-\frac{JS}{\sqrt{N}}\widehat{G}_{\textbf{k}\uparrow}(E)\nonumber
\end{equation}
\indent Let us now define the effective Green function,
\begin{equation}
\widehat{G}_{\uparrow}^{eff}(E-\frac{JS}{2})=\frac{1}{N}\sum_{\textbf{q}}[(E-\frac{JS}{2})\widehat{I}-\widehat{\epsilon}(\textbf{k-q})]^{-1}\nonumber
\end{equation}
\indent which upon substituting in the matrix form of the spin-flip equation of motion implies,
\begin{equation}\label{eq:spinflip-mp}
\frac{1}{\sqrt{N}}\sum_{\textbf{q}}\widehat{F}_{\textbf{k},\textbf{k-q},\textbf{q}\uparrow}(E)
= -JS\widehat{G}_{\uparrow}^{eff}(E-\frac{JS}{2})[\widehat{I}+\frac{J}{2}\widehat{G}_{\uparrow}^{eff}(E-\frac{JS}{2})]^{-1}\widehat{G}_{\textbf{k}\uparrow}(E)
\end{equation}
\indent Now , if we consider the Fourier transform of
eq.~\eqref{eq:Eqn-mot} and substitute eq.~\eqref{eq:spinflip-mp}
and eq.~\eqref{eq:Ising-mp} in it , we get the result for the
Green function as
%%\begin{widetext}
\begin{equation}
[(E+\frac{JS}{2})\widehat{I}-\widehat{\epsilon}({\textbf{k}})]\widehat{G}_{\textbf{k}\uparrow}(E)=\hbar\widehat{I}
+\frac{J^{2}S}{2}\widehat{G}_{\uparrow}^{eff}(E-\frac{JS}{2})[\widehat{I}+\frac{J}{2}\widehat{G}_{\uparrow}^{eff}(E-\frac{JS}{2})]^{-1}\widehat{G}_{\textbf{k}\uparrow}(E)\nonumber
\end{equation}
%%\end{widetext}
\indent i.e,
%%\begin{widetext}
\begin{equation}\label{eq:Green-matrix-fss-sd}
\widehat{G}_{\textbf{k}\uparrow}(E)=\hbar\widehat{I}[(E+i0^{+})\widehat{I}-\widehat{\epsilon}(\textbf{k})+\frac{JS}{2}\widehat{I}-\frac{J^{2}S}{2}\widehat{G}_{\uparrow}^{eff}(E-\frac{JS}{2})[\widehat{I}+\frac{J}{2}\widehat{G}_{\uparrow}^{eff}(E-\frac{JS}{2})]^{-1}]^{-1}
\end{equation}
%%\end{widetext}
\indent On comparing eq.~\eqref{eq:Green-matrix} and
eq.~\eqref{eq:Green-matrix-fss-sd} we get the self energy ansatz as
\begin{equation}\label{eq:selfenergymp-up}
\widehat{\Sigma}_{\uparrow}(E)=-\frac{JS}{2}\widehat{I}+\frac{J^{2}S}{2}\widehat{G}_{\uparrow}^{eff}(E-\frac{JS}{2})[\widehat{I}+\frac{J}{2}\widehat{G}_{\uparrow}^{eff}(E-\frac{JS}{2})]^{-1}
\end{equation}
\indent b) For $\sigma$ = $\downarrow$, the spin-flip function
vanishes;
\begin{eqnarray*}
F_{lm\sigma}^{\mu\nu}(E)=\langle\langle S_{l}^{-\sigma}c_{l\mu-\sigma};c_{m\nu\sigma}^\dagger \rangle\rangle_{E} =0 \nonumber
\end{eqnarray*}
\indent The equation of motion for the Green function, i.e,
eq.~\eqref{eq:Eqn-mot} reduces to,
\begin{eqnarray*}
(E-\frac{JS}{2})G_{lm\downarrow}^{\mu\nu}(E) =\hbar \delta_{lm}
\delta_{\mu\nu} + \sum_{p\gamma}
T_{lp}^{\mu\gamma}G_{pm\downarrow}^{\gamma\nu}(E) \nonumber
\end{eqnarray*}
\indent And upon taking Fourier transform we have,
\begin{eqnarray*}
(E-\frac{JS}{2})G_{\textbf{k}\downarrow}^{\mu\nu}(E) =\hbar
\delta_{\mu\nu} + \sum_{\gamma}
\epsilon^{\mu\gamma}(\textbf{k})G_{\textbf{k}\downarrow}^{\gamma\nu}(E)
\nonumber
\end{eqnarray*}
\indent which can be written in matrix form
\begin{equation}\label{eq:Green-matrix-fss-su}
\widehat{G}_{\textbf{k}\downarrow}(E)=\hbar\widehat{I}[(E-\frac{JS}{2}+i0^{+})\widehat{I}-\widehat{\epsilon}(\textbf{k})]^{-1}
\end{equation}
\indent Thus, the self energy becomes;
\begin{equation}\label{eq:selfenergymp-down}
\widehat{\Sigma}_{\downarrow}(E)=\frac{JS}{2}\widehat{I} \\
\end{equation}
\indent As seen, eq.~\eqref{eq:selfenergymp-up} and
eq.~\eqref{eq:selfenergymp-down} give the spin-dependent form of
the self energy which for the case of fully occupied bands and
ferromagnetically saturated spin sub-system can be written down as
follows;
%%\begin{widetext}
\begin{equation}
\widehat{\Sigma}_{\sigma}(E)=-\frac{JSz_{\sigma}}{2}\widehat{I}+\frac{J^{2}S(z_{\sigma}+1)}{4}\widehat{G}_{\sigma}^{eff}(E-\frac{JSz_{\sigma}}{2})[\widehat{I}+\frac{J}{2}\widehat{G}_{\sigma}^{eff}(E-\frac{JSz_{\sigma}}{2})]^{-1}
\end{equation}
%%\end{widetext}

\end{document}